\documentclass[structabstract]{aa}

\usepackage{graphicx}
\usepackage{amssymb}

\usepackage{natbib}

\usepackage{txfonts}

\begin{document}

\title{Extending the Hubble diagram by gamma ray bursts\\
\emph{(Research Note)}}
\titlerunning{Equation of State by Gamma Ray Bursts}
\authorrunning{L. Izzo. et al}
\author{L. Izzo\inst{1,2,3}, S. Capozziello\inst{1}, G. Covone\inst{1}
\and M. Capaccioli\inst{1,4}}

\institute{Dipartimento di Scienze Fisiche, Universit\`a di Napoli
"Federico II" and INFN Sez. di Napoli, Compl. Univ. Monte S.
Angelo, Ed. N, Via Cinthia, I-80126 Napoli, Italy, \and ICRANet
and ICRA, Piazzale della Repubblica 10, I-65122 Pescara, Italy,
\and Dip. di Fisica, Universit\`a di Roma "La Sapienza", Piazzale
Aldo Moro 5, I-00185 Roma, Italy,  \and INAF - VSTceN, Salita
Moiariello, 16, I-80131, Napoli, Italy}

\abstract {}{A new method to constrain the cosmological equation
of state is proposed by using combined samples of gamma-ray bursts
(GRBs) and supernovae (SNeIa). }{The Chevallier-Polarski-Linder
parameterization is adopted for the equation of state in order to
find out  a realistic approach to achieve the
deceleration/acceleration transition phase of dark energy
models.}{We find that GRBs, calibrated by SNeIa, could
be  good distance indicators capable of discriminating
between cosmological models and $\Lambda$CDM model at high redshift.}{}


\maketitle

\section{Introduction}
From an observational viewpoint, one of the fundamental goals
of cosmology is  to measure cosmological distances and then to build
up a suitable and reliable cosmic distance ladder. This issue has
recently become even more important due to the evident degeneracy of several dark
energy models with $\Lambda$CDM, despite the advent of the so-called
Precision cosmology, \citep{Ellis}.

In the last two  decades, a class of accurate
standard candles, the supernovae type Ia (SNeIa), has been highly
studied and the results obtained from the use of these objects led
to the surprising discovery of the  acceleration of the
cosmic Hubble flow  (for a review see Kowalski et al 2008). However these
objects are poorly detectable at redshifts higher than $\sim 1.5$,
so we need distance indicators at higher redshifts in order to
remove the  degeneration of dark energy models 
affecting  current cosmological models ($\Lambda$CDM is a good
approximation of the observed Universe, even though there is still no
theoretical basis about the nature of its components, but the issue
of global evolution is far from being addressed; for a comprehensive review see
Copeland et al 2006). A possible solution could be
found by adopting gamma ray bursts (GRBs) as distance indicators.

GRBs are  the most powerful explosions in the
Universe: the most likely scenarios for their generation are the
formation of  massive black holes or the  coalescence of binary
stellar systems. These events are observed at considerable
distances, so there have been several efforts to frame them into the
standard of the cosmological distance ladder. In the literature, there are
several models that  account for  GRB formation \citep{Meszaros}. All  these scenarios involve a similar shock
phenomenon: a "fireball",  possibly be supported by a further jet emission. 
 However none of these models
is intrinsically capable of integrating all the observable
quantities.

Despite  the poor knowledge of the GRB mechanism, it seems that
GRBs could be used as reliable  distance indicators. There
exist several observational correlations  among the photometric
and spectral properties of GRBs to support this possibility
\citep{Leandros, Ghirl2}. Nevertheless the origin of
these spectroscopic and photometrical correlations is not known very well
and there are several efforts to interpret
the behavior of GRB features in a coherent way, by  relatively
simple scenarios  \citep{MG, ghisellini}.
Succeeding in explain the mechanism that generates  GRBs is one of the
objectives of  modern astrophysics and to clarify these observed
correlations in this context would make GRBs  reliable distance indicators.
A complete review of the existing luminosity relations for GRBs can be found
in Schaefer (2007).

In this paper, we consider two relations, the one by
Liang-Zhang (LZ) \citep{LZ}, and the one by Ghirlanda (GGL)
\citep{GGL}. They are the only 3-parameter relations  known and  have
less scatter with respect to the theoretical best fit than the
other 2-parameter ones. However 
calibration of the relations used has been necessary in order to
avoid the circularity problem. This means that all the relations
need to be calibrated for each set of cosmological parameters.
Indeed, all GRB distances, obtained in a photometric way, are
strictly dependent on the cosmological parameters since,
currently, there is no low-redshift ($z$ up to 0.2-0.3) set of
GRBs available to achieve a cosmology-independent calibration. In order to
overcome this difficulty, Liang et al. (2008), proposed a
method in which several GRB relations have been calibrated by
SNeIa. Supposing that these relations work at all
redshifts and that, at the same redshift, GRBs and SNeIa have the
same luminosity distance, it becomes possible, in principle, to
calibrate the GRB relations using an interpolation algorithm. In
this way,  it becomes possible to  build a GRB-Hubble diagram by
calculating the luminosity distance for each GRB with the
well-known relation between the luminosity distance $d_l$ and the
energy-flux ratio of the distance indicators.

In the literature there are several paper that use similar methods to constraint
the cosmological parameters of the Concordance Model using GRBs as extension of the SNeIa Hubble Diagram,
\citep{Firmani1,Firmani2,Wang1,Wang2,Li1,Li}.

Here we take into account a cosmological EoS working at any redshift, using GRBs as tracers and adopting again
the Chevallier-Polarski-Linder  (CPL) parameterization. In particular we discuss a method which
should allow us to obtain an analytic cosmology-independent formulation of the
luminosity distance and then of the distance modulus. After a brief introduction to the GRB luminosity relations, we show fits of the data
obtained by these relations and the results and perspectives of the
approach are discussed in the last Section.

\section{The theoretical framework}

Our goal is to obtain an analytic formulation of the Hubble
diagram valid at any redshift. We start from
the Friedmann equation

\begin{equation}\label{eq:nostart}
 H^2 = \frac{8 \pi G}{3} \rho - \frac{k c^2}{a^2}\,.
\end{equation}
 We obtain, by some algebra,  the following equation in terms of
 the density parameter

\begin{equation}
 H^2 = H_0^2 \left[\Omega_0\left(\frac{a_0}{a}\right)^{3(w+1)} -
 \left(\Omega_0 - 1 \right)\left(\frac{a_0}{a}\right)^2\right],
\end{equation}
where the subscript $0$ indicates the present value of the
parameters. From now onwards, we take into account a spatially
quasi-flat Universe, $k \approx 0$; the contribution of the
curvature will be negligible and we have $\Omega_0 \approx 1$, as
suggested by the latest CMBR \citep{WMAP} and  the SNeIa
observations \citep{SCP}. However, in the final section, we will perform
a test to verify this assumption with  observations coming from
GRBs. Now if we translate in terms of redshift $z$,

\begin{equation}
 \frac{a_0}{a} = 1+z\,,
\end{equation}
the previous equation reduces to

\begin{equation}\label{eq:no1}
 H^2(z) = H_0^2 \left( 1 + z\right)^{3(w+1)}\,.
\end{equation}
The $w$-parameter indicates  the EoS $w = p/\rho$, where $p$ and
$\rho$ are  the pressure and the matter-energy density of the
Universe, respectively. Considering the CPL parameterization of
the EoS, \citep{CPL}:

\begin{equation}
 w(z) = w_0 + w_a \frac{z}{1+z},
\end{equation}
and substituting  into Eq.(\ref{eq:no1}), we obtain:

\begin{equation}
 H(z) = H_0 \left[\left(1+z\right)^{\frac{3}{2}(w_0 + w_a + 1)} \exp{\left(\frac{-3w_a z}{2(1+z)}\right)}\right],
\end{equation}
which enters directly in the expression of the distance modulus

\begin{equation}\label{eq:no2}
\mu(z) = -5 + 5 \log{d_l(z)}\,,
\end{equation}
where $ d_l(z) = c (1+z) D_l(z)$ and where
\begin{equation}\label{eq:no2a}
D_l(z)=\int_0^z \frac{d \xi}{H(\xi)}\,. \end{equation} This means
that an analytic expression for $\mu$ can be achieved. The
integral $D_l$ in Eq.(\ref{eq:no2a}) can be solved giving a Gamma
function of the first kind \footnote[1]{In our  case, the variable
of the Gamma function, $z$, is always positive so that we have no
problem of discontinuity in applying the gamma function in the
following calculations.}:

\begin{eqnarray}\label{eq:Luca}
 D_l(z) =  \left(\frac{3 w_a}{2}\right)^{-\frac{1+3 w_0+3 w_a}{2}}
\nonumber
\\
\left. \exp{\left(\frac{3 w_a}{2}\right)} \Gamma \left[ \frac{1+ 3 w_0 + 3 w_a}{2}, \frac{3 w_a}{2 (1+\xi)}\right]
\right|_{\xi = 0}^{\xi = z}.
\end{eqnarray}

Substituting  such an expression in the distance modulus, we
obtain a model for  data fitting which could work, in principle,
at any $z$. The obtained expression
for the Hubble parameter $H(z)$ is independent of the density
parameters, $\Omega_M$ and $\Omega_{\Lambda}$, provided that their sum is equal to 1.

We  use the CPL parameterization
not only for the dark energy component, but  for the total
energy-matter density of the Universe. This assumption works
because dark and baryonic matter  contribute with a null
pressure while the radiation component is negligible in matter-
and dark energy-dominated eras. Furthermore, the analytical
formulation that we  adopt  for the luminosity distance is
assumed valid at any redshift $z$.

\section{GRB luminosity relations}

In the last years, thanks to several spacecraft missions capable of observing this high energy
region, the main features of GRBs have become better known.
Recently, some photometric and spectroscopic relations between GRB
observables have been found and then the hypothesis that these
objects could be considered  suitable distance indicators has become feasible.
Nevertheless,  there is no
theoretical model that fully explains these relations so the GRBs
cannot be considered as standard candles. For a
detailed review of the observational features see
Schaefer (2007).

Here, we  take into account the existing 3-parameter
relations. This choice has been made because these relations place
 better constraints on the data giving less scatter between the
theoretical relation and the experimental data 
\citep{Schaefer}. The first relation is the so-called Liang-Zhang
relation, \citep{LZ}, which allows us to connect the GRB peak energy,
$E_p$, with the isotropic energy released in the burst, $E_{iso}$,
and with the  jet break - time of the afterglow optical light
curve in the rest frame, measured in days, $t_b$, that is
\begin{equation}\label{eq:noLZ}
 \log{E_{iso}}=a + b_1 \log{\frac{E_p (1+z)}{300keV}} + b_2 \log{\frac{t_b}{(1+z)1day}}
\end{equation}
where $a$ and  $b_i$, with $i=1,2$, are calibration constants.

The other  relation is that  given by Ghirlanda et
al \citep{GGL}. It connects the peak energy $E_p$ with the
collimation-corrected energy, or the energy release of a GRB jet,
$E_{\gamma}$, where $E_{\gamma} = F_{beam} E_{iso} = 1 - \cos({\theta}) E_{iso}$, with $\theta_{jet}$ the jet
opening angle defined in  Sari et al (1999):

\begin{equation}
 \theta_{jet} = 0.163\left(\frac{t_b}{1 + z}\right)^{3/8}\left(\frac{n_0\eta_{\gamma}}{E_{iso,52}}\right)^{1/8},
\end{equation}
where $E_{iso,52} = E_{iso}/10^{52}$ ergs,  $n_0$ is the
circumburst particle density in 1 cm$^{-3}$, and $\eta_{\gamma}$
the radiative efficiency. The Ghirlanda et al. relation is

\begin{equation}\label{eq:noGGL}
 \log{E_{\gamma}} = a + b \log{\frac{E_p}{300 keV}},
\end{equation}
where $a$ and $b$ are two calibration constants.

From these relations, we can  directly obtain the luminosity
distance $d_l$ from the well-known formula which connects $d_l$
with the isotropic energy $E_{iso}$ and the bolometric fluence
$S_{bolo}$ :
\begin{equation}
 d_l = \left(\frac{E_{iso}}{4 \pi S_{bolo}}\right)^{\frac{1}{2}},
\end{equation}
from which it is easy to compute, for each GRB, the distance
modulus $\mu = $ and its error given by \citep{Liang}:
\begin{equation}
 \sigma_{\mu} =  \left[\left(2.5 \sigma_{\log{E_{iso}}}\right)^2  + \left(1.086\sigma_{S_{bolo}}/S_{bolo}\right)^2\right]^{\frac{1}{2}}
\end{equation}
with $\sigma_{\log{E_{iso}}}$ and $\sigma_{S_{bolo}}$  obtained
from the error propagation applied to Eq.(\ref{eq:noLZ}) and
Eq.(\ref{eq:noGGL}).  Moreover, we assume that the error in the
determination of the redshift $z$ is negligible, as well as for
the radiative efficiency $\eta_{\gamma}$. We note also that the
assumption of a well-known $n_0$ is a strong hypothesis since the
goodness of the fits depends, in particular, on this parameter.
The  GRB data sample  is taken from the already cited work by Schaefer. We  take into account 27 events with
extremely precise data. This sample is the same one adopted in
Capozziello $\&$ Izzo (2008).

\section{The data fitting}

The next step is the  fit of the GRB sample with the empirical
relations, Eqs.(\ref{eq:noLZ}),(\ref{eq:noGGL}), described in
Sect. 3. The aim is to achieve an estimate of the CPL parameters
and consequently to determine the trend of the EoS at any
redshift, using the analytical relation, Eq.(\ref{eq:Luca}). We are considering the same sample of 27 GRBs used in
Capozziello $\&$ Izzo (2008)  in which we have added the
sample of SNeIa  by the Union Supernova Survey  \citep{SCP}.

The numerical results of the fits are shown in Table
\ref{table:no1}, where we obtain a robust estimation of the
CPL parameters for both the relations used, with and without SNeIa
data.  An immediate comparison is done with the best fit applied
only to the SNeIa sample. It is evident how adding GRBs to SNeIa
data completes the knowledge of and the accuracy on the EoS parameter
$w$.

\begin{table}
\caption{Results of the fits. SNeIa is only for the supernovae  data,
LZ is for the GRBs data obtained from the Liang-Zhang relation,
GGL for the Ghirlanda et al. one. } % title of Table
\label{table:no1} % is used to refer this table in the text
\centering % used for centering table
\begin{tabular}{l c c c} % centered columns (4 columns)
\hline\hline % inserts double horizontal lines
Relation & $w_0$ & $w_a$ & $R^2$ \\ % table heading
\hline % inserts single horizontal line
 SNeIa & $-0.910 \pm 0.070$ & $0.755 \pm 0.054$ & $0.983$ \\
 LZ & $-1.39 \pm 0.38$ & $1.18 \pm 0.37$ & $0.817$ \\
 GGL & $-1.46 \pm 0.38$ & $1.36 \pm 0.32$ & $0.812$ \\
 LZ + SNeIa & $-1.15 \pm 0.10$ & $0.93 \pm 0.11$ & $0.933$ \\
 GGL + SNeIa & $-1.42 \pm 0.12$ & $1.24 \pm 0.13$ & $0.920$ \\
\hline %inserts single line
\end{tabular}
\end{table}

\begin{center}
\begin{figure}
\includegraphics[width=9cm, height=6 cm]{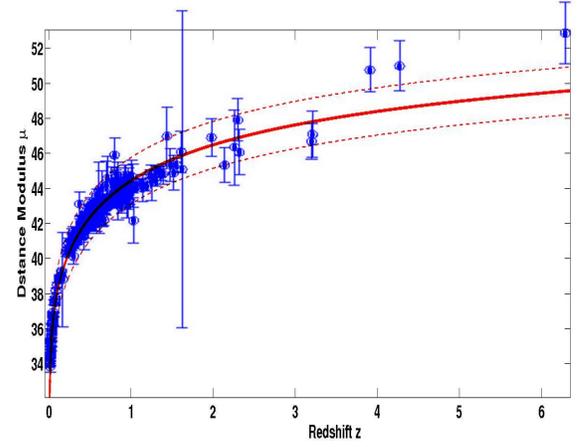}
\caption{Redshift-distance modulus diagram for the GRB+SNeIa
sample. The black dots are the GRBs , the blue ones are the SNeIa.
The red line is the best fit obtained from the data, with the
dashed line representing the confidence limits at $3\sigma$. The
error bars on the supernova data are not represented because they
are negligible.}
\label{fig:no1}
\end{figure}
\end{center}

In order to measure the goodness of the fit, we use the $R^2$
test, see Table \ref{table:no1}. The
$R^2$ test is a measure of how successful the fit is in
explaining the variation of the data (see for details
Draper $\&$ Smith 1998). An $R^2$ close to 1.0 indicates that we have accounted for almost all of the variability with the data specified in the model.
 As a  standard, the $R^2$ test is the square of
the correlation between the response values and the predicted
response values, that is:
\begin{equation}
 R^2 = 1 - \frac{SSE}{SST} = 1 - \frac{\sum_{i = 1}^{n} w_i (y_i - \hat{y}_i)}{\sum_{i = 1}^{n} w_i (y_i - \bar{y}_i)^2},
\end{equation}
where $SSE$  is the sum of the squares due to errors and it
measures the total deviation of the response values from the fit and
 $SST$ is the sum of squares about the mean: $\hat{y}$ is the predicted response value, $\bar{y}$ is the
mean value and the $w_i$ are the weights on the values.

\begin{center}
\begin{figure}
\includegraphics[width=9cm, height=6.5 cm]{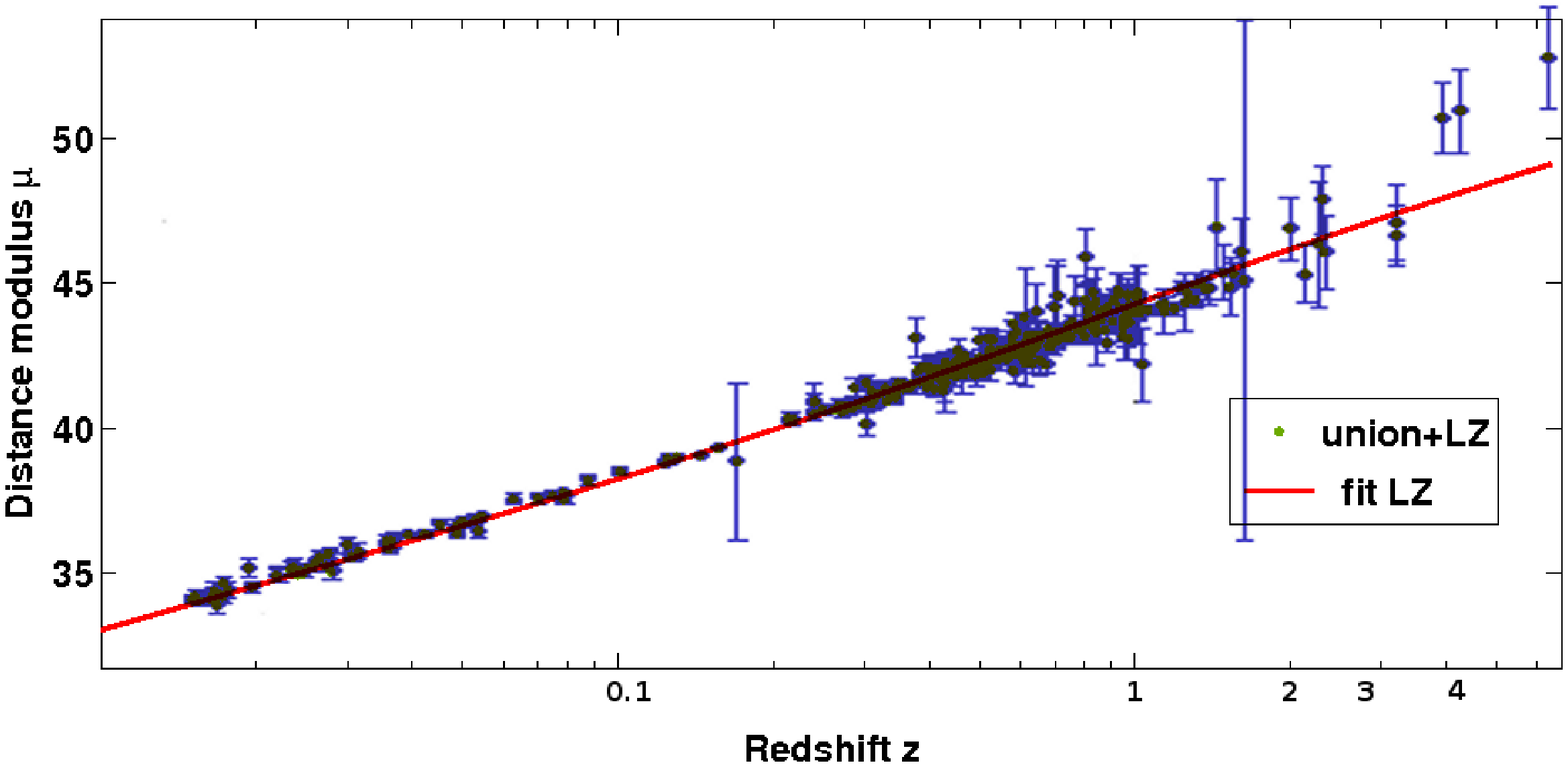}
\caption{Redshift-distance modulus diagram for the GRB+SNeIa
sample versus redshift in logarithmic scale. }
\label{fig:no5}
\end{figure}
\end{center}

The extension of the supernova Hubble Diagram with the GRB data can
be used to improve our knowledge of the trend at high redshift. In this way, also using
the GRB data, we show in Fig. ~\ref{fig:no5}, the
distance modulus $\mu$ versus the redshift $z$, in a logarithmic
scale. The best fit curve, obtained with  Eq.(\ref{eq:Luca}), is also reported. A more detailed analysis confirms the
presence of a transition (re-acceleration) redshift around $z=0.5$.

\begin{figure}
\includegraphics[width=8.5 cm, height=5 cm]{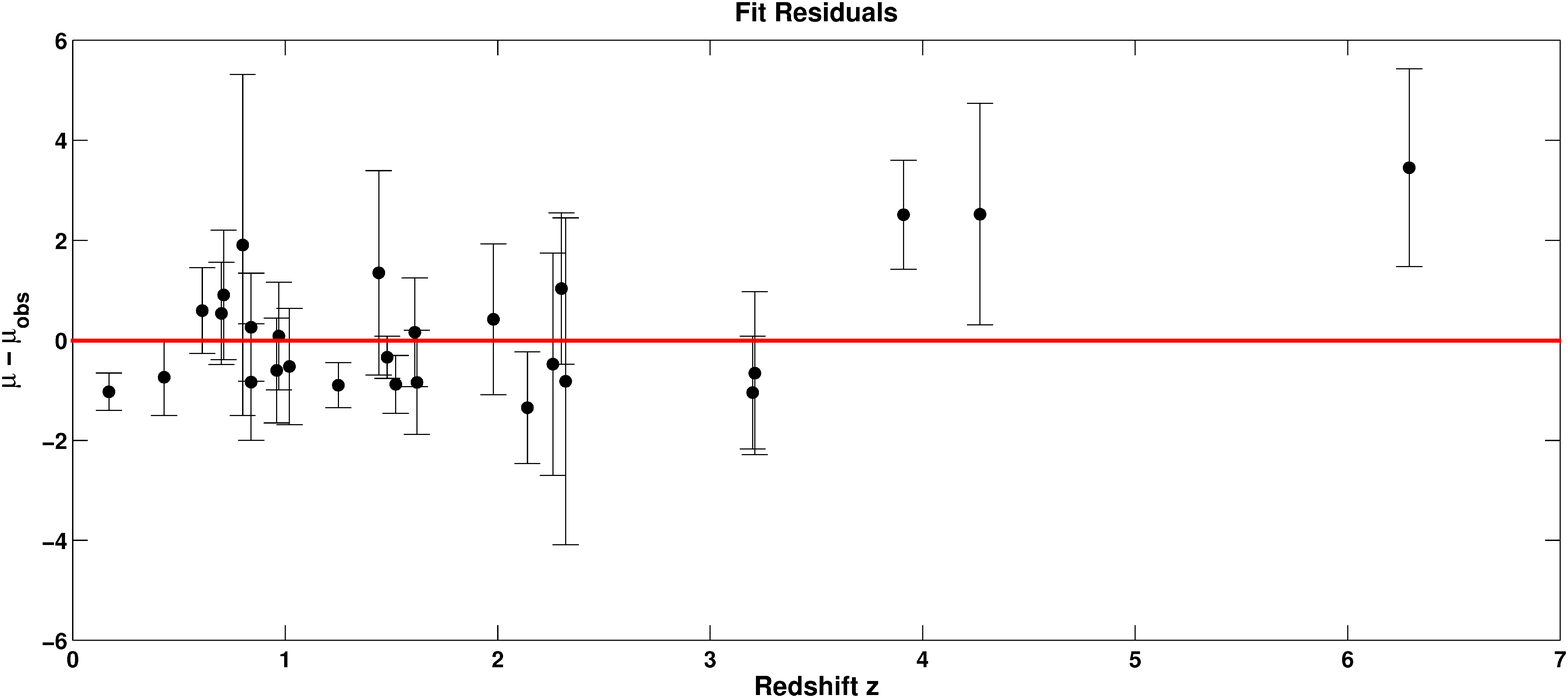}
\caption{Comparison between  the best fit of $\mu$ and the
observed distance modulus $\mu_{obs}$ at any redshift. The black
dots are the GRBs data and the red line is the best fit curve
representing the theoretical distance modulus.}
\label{fig:no3}
\end{figure}

In Fig. ~\ref{fig:no3}, we plot the comparison between the
theoretical $\mu_{th}$ and the observed distance modulus
$\mu_{obs}$ at any redshift, the residual plot. A smooth trend up
to $ z \approx 3.5$ in the residual curve can be immediately
detected.  Beyond this limit, we have 3 GRBs that exceed, by the
same side, the $3\sigma$ confidence limit of the best fit. This
discrepancy is clear in Fig. ~\ref{fig:no5}, where  we plot  the
best fit for the combined sample in the case of an LZ relation with a
logarithmic scale for the redshift.

This fact is fundamental  for the goodness of the fit because
these GRBs represent the most distant objects that one can use to
make such an analysis and their weight on the fit is very high, in the sense that they appear to be not accurate
distance indicators.

However there is strong evidence that the data of these 3 GRBs,  GRB 050505, GRB 050904 and GRB 060210, reported in the Schaefer catalog, are uncertain.
For the first and the last GRB the peak energies are underestimated \citep{Cabrera}. As a consequence we would obtain an underestimated value for the bolometric fluence.
GRB 050904 is the most distant GRB considered and it shows some probelms related to the peak energy reported by different authors, that differ by a factor of $\sim 3$.
Moreover the afterglow of GRB is complicated by flares and re-brightenings so that the standard afterglow model
gives an extremely high value of the circumburst medium density $n \sim 700
cm^{-3}$  \citep{Frail}, contrary to the assumed value $n = 3 cm^{-2}$.

\begin{center}
\begin{figure}
\includegraphics[width=9 cm, height=6.5 cm]{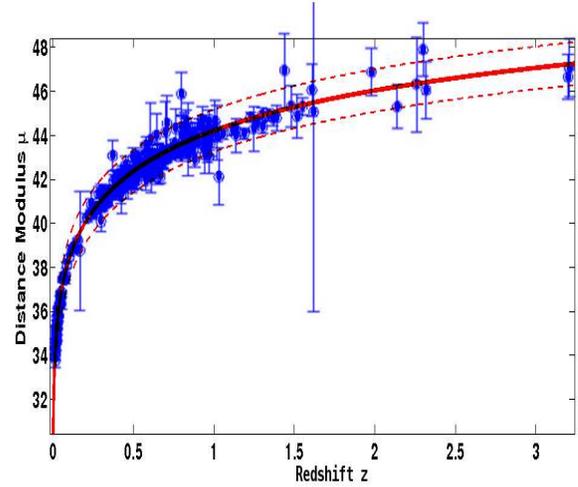}
\caption{Redshift-distance modulus diagram for the corrected GRB+SNeIa
sample. The red line is the best fit obtained from the data, with the
dashed line representing the confidence limit at $3\sigma$.}
\label{fig:no7}
\end{figure}
\end{center}

For this reason we repeat the analysis described above without these 3
GRBs, obtaining a better value than the previous one for the
$R^2$ test. The results of these corrected fits are shown in
Table \ref{table:no3}. In Fig. ~\ref{fig:no7},  we plot
the best fit with this corrected sample. From these results we
conclude that the complete sample gives different results than does
corrected sample, the first one suggesting a phantom/quintessence
regime for the present epoch while the second one 
fits an accelerating $\Lambda$CDM model. This last
result is confirmed by the following analysis, where we have
performed a Monte-Carlo-like procedure for the comparison of the
results with the usual likelihood estimator given by
\begin{equation}
\chi^2 = \sum_{i = 1}^{N} \left[\frac{(\mu_{th}(z_i) - \mu_{obs}(z_i))^2}{\sigma_i}\right],
\end{equation}
in the context of a $\Lambda$CDM model of the Universe,  where
$\mu_{th}$ is the distance modulus computed from the
Eq.(\ref{eq:no2}) and Eq.(\ref{eq:no2a}), $z_i$ is the  observed
redshift for each GRB and $\sigma_i$ the observed distance modulus
uncertainty. The results of this analysis are shown in  Table
\ref{table:no4}, where we can see the improvement obtained by the
GRB sample corrected for the 3 "wrong" GRBs.

\begin{center}
\begin{figure}[ht]
\includegraphics[width=8.5cm, height=6.5 cm]{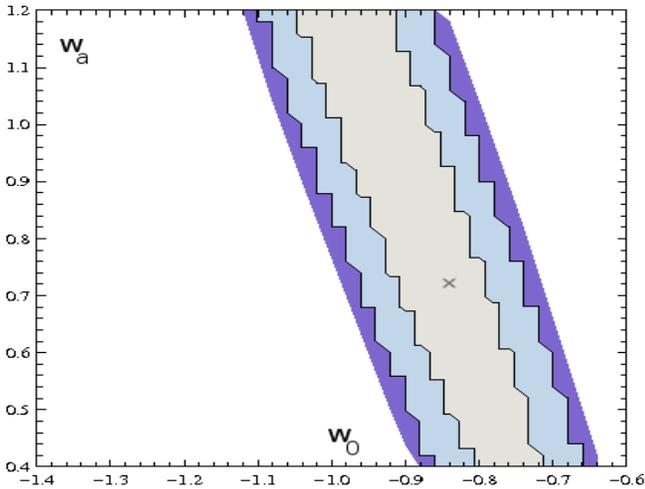}
\caption{68\%, 95\% and 98\% constraints on $w_0$  and $w_a$
obtained from the UNION sample and the GRB sample corrected for the 3
outlying GRBs. The cross represents the best fit value and it is in a
good agreement with that found using the theoretical model
described in Sect.2.} \label{fig:no9}
\end{figure}
\end{center}

We have adopted a similar procedure in the case of an EoS evolving
with redshift and where $\mu_{th}$ is obtained by 
Eq.(\ref{eq:Luca}). The result of this analysis is plotted in Fig.
~\ref{fig:no9} where the best fit value, the cross in the figure,
corresponds to the value $w_0 = -0.84 \pm 0.14$ and $w_a = 0.72
\pm 0.06$, and  where the boundaries correspond to 1$\sigma$, 2$\sigma$
and 3$\sigma$ confidence levels, in a good agreement with the results obtained, see
Table \ref{table:no3}, using our theoretical relation,
Eq.(\ref{eq:Luca}).

From this analysis, we conclude that the corrected
sample agrees fairly well with the $\Lambda$CDM model with a small
contribution of the curvature parameter,  $k = 0.01
\pm 0.04$. Thus, the method delineated in  Sect.2
seems a good approximation of the observed cosmography and agrees
very well with  the $\Lambda$CDM model, so that we can argue  that
GRBs could be good distance indicators at redshift values up to $z
= 4$.

\begin{table*}[ht]
\caption{Cosmological density parameters, with uncertainties
computed at 1$\sigma$ confidence limit, obtained by a Monte Carlo procedure} % title of Table
\label{table:no4} % is used to refer this table in the text
\centering % used for centering table
\begin{tabular}{l c c c c} % centered columns (4 columns)
\hline\hline % inserts double horizontal lines
Sample & $\Omega_m$ & $\Omega_{\Lambda}$ & $\Omega_k$ & $\chi^2$ \\ % table heading
\hline % inserts single horizontal line
 UNION + GRB & $0.26 \pm 0.14$ & $0.73 \pm 0.14$ & $0.01 \pm 0.04$ & $1.032$ \\
 UNION + GRB corrected & $0.25 \pm 0.10$ & $0.74 \pm 0.135 $ & $0.01 \pm 0.035$ & $1.00027$ \\ % inserting body of the table
\hline %inserts single line
\end{tabular}
\end{table*}


\section{Discussion and conclusions}

Starting from the Friedmann equation, we have investigated a new
method to constrain the cosmological equation of state at high
redshifts.  In particular we obtain an analytical formula for the distance modulus
so we could directly estimate  the parameters of the cosmological model considered.
The working hypothesis involves the use of GRBs as
distance indicators  at  high redshift, well beyond the distance
where SNeIa have been detected  to date.  The CPL parameterization for
the EoS has been explicitly used for the whole matter-energy
content of the Universe as a suitable approach to investigate the
parameter $w=w(z)$ and discriminate values with respect to the
$\Lambda$CDM model. In particular, regarding the Friedmann
equations, we have obtained, in the case of the LZ relation, the epoch for the transition between the
deceleration-acceleration phases at a redshift value of $z \approx
5 $,  with a reliable confidence level. This is a value that,  if higher than the redshift of the
farthest GRB used, could be in agreement with current
quasar formation scenarios. Also, we are in good agreement with
the observed phantom/quintessence regime at the present epoch, that is
for $z \rightarrow 0$, we obtain $w \leq -1$.

So we reject the current phantom
regime by this analysis, obtaining for $w_0$ a value  in
agreement with the $\Lambda$CDM model at the present epoch. The
method, while preliminary, seems to indicate that GRBs could be
used as standard candles once a reliable unified model
of their  photometric and spectroscopic quantities is achieved
(some relevant results  are presented in
Ghisellini et al 2008). However, more robust samples of data are
needed and a more realistic EoS  (with respect to the simple perfect
fluid models) should be taken into account in order to suitably
track  redshift at any epoch (see for example Capozziello et al 2006).

With  improving observations, in particular with the
launch of new satellites devoted to  GRB surveys, such as
Fermi-GLAST\footnote[3]{http://fermi.gsfc.nasa.gov} and
AGILE\footnote[4]{http://agile.rm.iasf.cnr.it}, one should be able
to expand the samples of GRBs, possibly with data coming from
objects at higher redshift.

Considering  these preliminary results, it seems that
GRBs could be considered as a useful tool to remove degeneracy
and constrain self-consistent cosmological models. The
matching with other distance indicators  would improve the
consistency of the Hubble distance-redshift  diagram by extending
it up to redshift $6\, -\, 7$ and higher.

\begin{table}
\caption{Results of the fits corrected for the 3 ``outying'' GRBs.
SNeIa is  for the supernova Ia data, LZ is for the GRBs data obtained from the Liang-Zhang relation,
GGL for the Ghirlanda et al. one.} % title of Table
\label{table:no3} % is used to refer this table in the text
\centering % used for centering table
\begin{tabular}{l c c c} % centered columns (4 columns)
\hline\hline % inserts double horizontal lines
Relation & $w_0$ & $w_a$ & $R^2$ \\ % table heading
\hline % inserts single horizontal line
 LZ + SNeIa & $-0.95 \pm 0.01$ & $0.74 \pm 0.01$ & $0.999$ \\
 GGL + SNeIa & $-0.865 \pm 0.005$ & $0.66 \pm 0.005$ & $0.999$ \\
\hline %inserts single line
\end{tabular}
\end{table}

\end{document}